\documentstyle[11pt,aas2pp4]{article}

\begin{document}

\title{\Large UV Spectral Dating of Stars and Galaxies$^{\dagger }$  }

\author{S.R. Heap$^{1}$, T.M. Brown$^{1,11}$, I. Hubeny$^{1,11}$, W. 
Landsman$^{1,3}$, 
 S. Yi$^{1,5}$,\\
M. Fanelli$^{1,3}$, J.P. Gardner$^{1,11}$, T. Lanz$^{1,4}$, S. P. Maran$^{2}$,
A. Sweigart$^{1}$,\\
M.E. Kaiser$^{1,10}$, J. Linsky$^{6}$, J.G. Timothy$^{7}$,\\
D. Lindler$^{1,8}$, T. Beck$^{1,8}$, R.C. Bohlin$^{9}$, M. Clampin$^{9}$,\\
J. Grady$^{1}$, J. Loiacono$^{1}$, C. Krebs$^{2}$ }

\noindent 
1. Laboratory for Astronomy \& Solar Physics, Code 681, NASA/GSFC, 
   Greenbelt MD 20771\newline
2. NASA/Goddard Space Flight Center, Greenbelt MD 20771\newline
3. Hughes STX, 440 Forbes Blvd., Lanham MD 20706\newline
4. Univ. of Utrecht, Utrecht, The Netherlands\newline
5. NAS/NRC Fellow\\
6. Univ. of Colorado, Boulder CO 80309-0440\newline
7. Univ. of New Brunswick, P.O. Box 4400, Fredricton NB, E3B 5A3, Canada\newline
8. Advanced Computer Concepts, Potomac MD 20854\newline
9. Space Telescope Science Institute, 3700 San Martin Drive, Baltimore MD 
   21218\\
10. Dep't of Physics \& Astronomy, Johns Hopkins Univ., Baltimore MD 21218\\
11. NOAO Research Associate\\

\noindent $\dagger $Based on observations with the NASA/ESA Hubble
Space Telescope, obtained at the Space Telescope Science Institute, which is
operated by the Association of Universities for Research in Astronomy, Inc.,
under NASA contract NAS5-2655.

\date{\today }

\begin{abstract}
An echelle spectrogram (R~=~30,000) of the 2300--3100~\AA\ region in
the ultraviolet spectrum of the F8V star 9 Comae is presented.  The
observation is used to calibrate features in the mid-ultraviolet 
spectra of similar stars according to age and metal content.  In 
particular, the spectral break at 2640~\AA\ is interpreted using 
the spectral synthesis code \textsc{synspec}.  We 
use this feature to estimate the time since the last major star
formation episode in the $z=1.55$ early-type galaxy LBDS 53W091, whose 
rest frame mid-ultraviolet spectrum, observed with the Keck Telescope,
is dominated by the flux from similar stars that are at or near 
the main-sequence turnoff in that system (Spinrad et al.\ 1997\markcite{S97}).
Our result, 1 Gyr if the flux-dominating stellar population has
a metallicity twice solar, or 2 Gyr for a more plausible solar metallicity, is 
significantly lower than the previous estimate and thereby relaxes
constraints on cosmological parameters that were implied by the 
earlier work.

\end{abstract}


\section{Introduction}

Elucidating the formation and evolution of galaxies is a prime goal of
the Hubble Space Telescope and other major telescopic facilities.  To 
accomplish this, it is important to obtain and interpret spectral 
information on distant galaxies that may be in early phases of evolution.
The very red radio galaxy, LBDS 53W091 at 
redshift $z=1.552$, recently 
examined spectroscopically in the rest frame ultraviolet
with the Keck Telescope 
(Dunlop et al.\ 1996\markcite{D96}; Spinrad et al.\ 1997\markcite{S97}), is
a potential milestone in this work.  These observations may constrain
the earliest epoch of star formation, and thereby, perhaps, cosmological
parameters.  To interpret such spectra, it is necessary to understand
what stellar population(s) contribute to them.  This can be approached 
by obtaining ultraviolet spectra of suitable quality of nearby stars
whose nature is well understood.

The Keck spectrum of galaxy 53W091 is remarkably similar to the rest frame
ultraviolet spectrum of a mid-to-late F star.  The
similarity is quite strong near 2600~\AA, although the spectrum becomes
a composite of later
types (late F to early G) near 2900~\AA.  It clearly shows the
principal absorption features of \ion{Mg}{1} and \ion{Mg}{2}, and the two 
well-known spectral breaks at $\lambda_{\rm rest} = 2640$ and $2900$~\AA.  
This marked resemblance of a galaxy spectrum to that of stars from a small
range in stellar types 
is due to the overwhelming contributions in the mid-ultraviolet spectrum of
intermediate-age stars near the main sequence
turnoff.  Since the temperature or spectral type at the turnoff is a prime
age indicator for a coeval population, the dominating fluxes of such stars
in the mid-ultraviolet have obvious application to dating unresolved
stellar systems.  

The Keck observers of 53W091 derived an age of at least 3.5 Gyr since 
the last epoch of major star formation in that system.  An age that 
great would rule out cosmologies with deceleration parameter 
$\Omega =1$  and Hubble constant H$_{\mathrm{o}} \geq 50$ 
km s$^{-1}$ Mpc$^{-1}$. Such a strong conclusion deserves further
test.  In particular, it is desirable to make an improved calibration of 
the mid-ultraviolet spectra of F-type stars, to support the application of
this dating method to this and other distant galaxies.  Fanelli et 
al.\ (1992\markcite{F92}) 
used low-resolution IUE spectra to show that the spectral
breaks at 2640 and 2900~\AA\ correlate with spectral type (temperature) but
are only mildly sensitive to metallicity.  To refine and extend this
calibration with the first-generation spectrographs on the Hubble Space
Telescope appeared impractical.  However, thanks to the multiplex 
advantages of an imaging spectrograph, 
we have initiated a program to this end with the Space 
Telescope Imaging Spectrograph (STIS).

Our STIS General Observer program \#7433 will obtain mid-ultraviolet
spectra of twelve F-type stars of known atmospheric properties at a
resolution $\lambda/\Delta \lambda = 30,000$. Here we
report representative observations of one such star, 9 Comae (metal rich, 
F8V).  With these results, we improve the mid-ultraviolet calibration
of this and related stars and thereby conclude that the
corresponding rest frame spectrum of 53W091 is compatible with a much younger
age ($\lesssim$ 2 Gyr) than reported by 
Dunlop et al.\ (1996\markcite{D96}) or Spinrad et al.\ (1997\markcite{S97}).  
This relaxes the strong constraints on cosmology that were 
implied by the earlier work.

\section{Observations \& Reduction}

The F8V star 9 Comae (HR 4688, HD 107213) is expected to resemble the stars
responsible for the rest frame mid-ultraviolet spectrum of 53W091.  Its 
atmospheric properties (Edvardsson et al.\ 1993\markcite{E93}) are well known:
T$_{\rm eff}$~=~6343~K, log $g$~=~4.05, 
[M/H]~=~+0.28 and [Fe/H]~=~+0.21.  The star lies at a Hipparcos distance
of 49.7 pc, well in front of 
the open cluster Mel 111.  The
age of 9 Comae is estimated at 1.6--2.4 Gyr depending on whether it is
on the ``blue hook'' in the HR diagram or still on 
the main sequence (c.f.\ Figure 1). Because this star is
metal-rich, its STIS spectrum provides a stringent test of the completeness
of the atomic and molecular line list used in its interpretation.
However, we note that 
the star is somewhat more metal-rich than the abundances expected
in the integrated spectrum of 53W091.  Many studies
(Gonz$\acute{\rm a}$lez \& Gorgas 1995\markcite{G95} and references therein) 
suggest that there is a radial metallicity gradient
in elliptical galaxies.  Since the Keck spectrum was observed
through an aperture that integrated out to the effective radius, 
the ``average'' metallicity in the Keck spectrum is probably close to solar.

\begin{figure}[!h]
\plotone{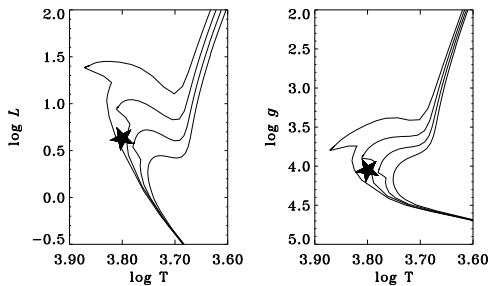}
\caption{
9 Comae (star symbol) on the HR diagram, 
plotted in the traditional way (left) 
and in the T$_{\rm eff}$-log $g$ plane (right). 
The four isochrones (Yi \& Demarque 1997) 
are for [M/H]~=~+0.3 and ages of 1, 2, 4, and 
8 Gyr (top to bottom). 
}
\end{figure}

The STIS ERO observing program \#7139 on 9 Comae was accomplished within a 
single 96-minute orbit of the HST. During this period,
the star was located with the STIS CCD detector, the pointing of the 
Telescope was adjusted to place the stellar image in the 
0\farcs2$\times$0\farcs063 
entrance slit, and the spectrum was recorded by the near-ultraviolet MAMA
detector.
The total integration was 20 minutes, divided into two sub-exposures of ten 
minutes each.  The resulting echellogram covering the
spectral region 2280--3110~\AA\ is shown in Figure 2 (Plate XX).

\begin{figure*}[p]
\plotfiddle{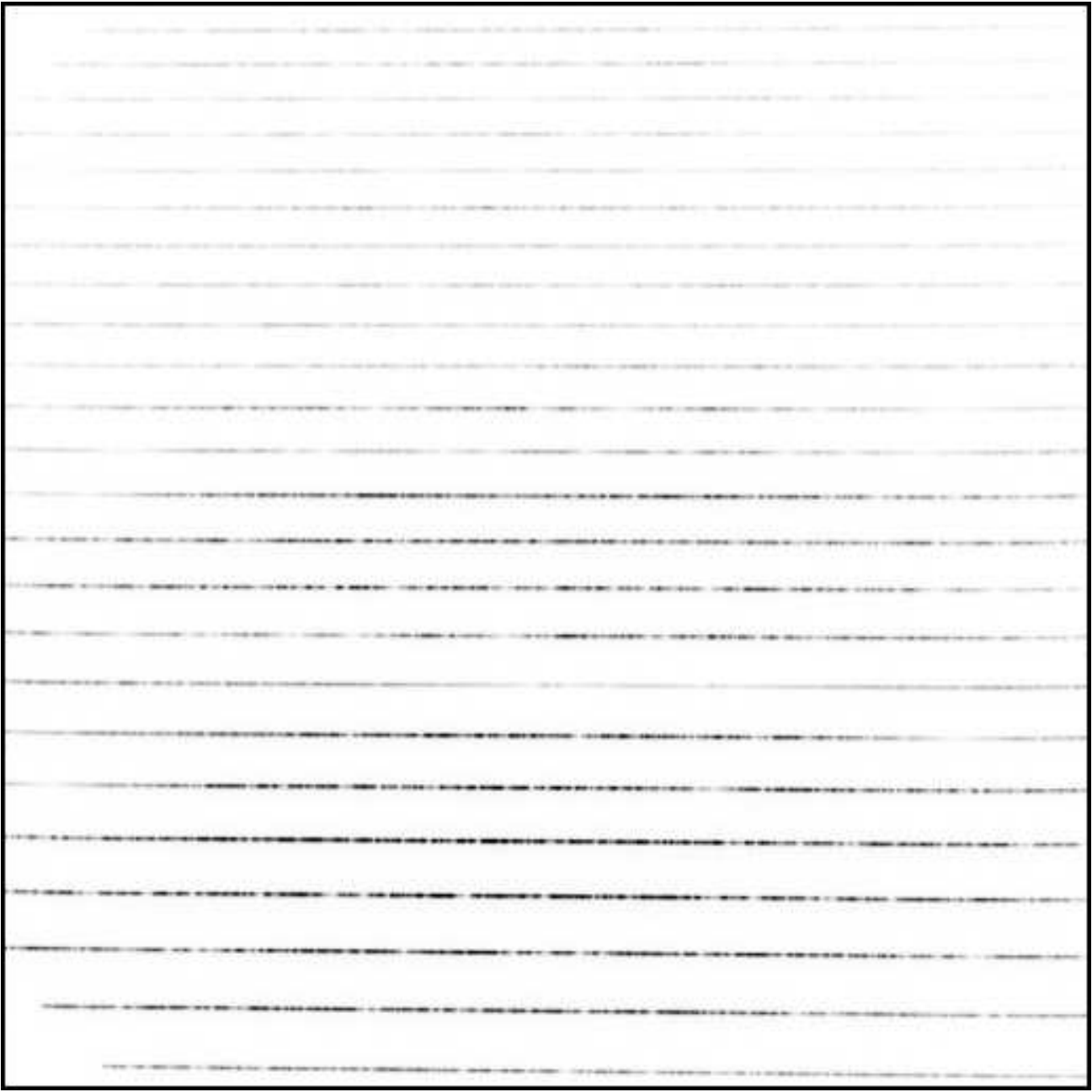}{6.5in}{0}{90}{90}{-275}{-110}
\caption{
STIS echellogram of 9 Comae. The gray-scale is reversed so that the
brightest regions are black. Wavelength increases to the right (within an
order) and down (from order to order). 
}
\end{figure*}

At the time of observation, 1997 May 31, the operating 
procedures to ``peak-up'' a target in a STIS entrance aperture were not 
yet optimized.  Accordingly, the fraction of the light in the stellar
image that was transmitted through the aperture was less than the nominal
value applicable when STIS is fully commissioned.  Nevertheless, the
signal-to-noise ratio in the continuum
is 40 per pixel above 2800~\AA, and 15 shortward of 2600~\AA.  Again,
because STIS was not fully operational at the time of observation, the
doppler compensation program that adjusts for the changing velocity 
vector of the spacecraft during the observation was not active. However,
any smearing due to spacecraft motion in this spectrum is less than the
rotational velocity of 9 Comae or the nominal 2-pixel 
resolution (10 km s$^{-1}$) of the spectrograph in this 
mode.

The STIS Investigation Definition Team version of the CALSTIS program
was used to reduce the spectrogram through the stage of
extracting the spectra from the various grating orders and subtracting 
the inter-order spectral background.  Next, we corrected for the echelle 
blaze function by Barker's (1984\markcite{B84}) method and merged the 
data from the 
various orders into a single spectrum comprising almost 20,000 data 
elements.  In this spectrum, saturated absorption lines reach only to
a residual
intensity $I=0.03$ rather than to zero.  This may be due to instrumental
scattered light associated with the echelle grating.  Alternatively,
it may result from chromospheric emission in the cores of the strong
lines (chromospheric emission is quite obvious in the cores of the
\ion{Mg}{2} $\lambda\lambda$2795,2803 doublet).

As an in-orbit radiometric calibration is not yet available for STIS 
mode E230M spectra, we calibrated the observation using component
sensitivities measured before launch by the STIS team.  Figure 3 shows 
the derived relative flux distribution of 9 Comae, binned to 1~\AA\ 
resolution.  Due to the uncertainty in calibration, we concentrate 
our analysis on the two spectral breaks B($\lambda 2640$) and B($\lambda 2900$)
that were used to date the time since the last major star-forming event in
53W091 (Spinrad et al.\ 1997\markcite{S97}).  
These two features, also shown in 
Figure 3, are insensitive to errors in flux calibration and interstellar
reddening.

\begin{figure}[h]
\plotone{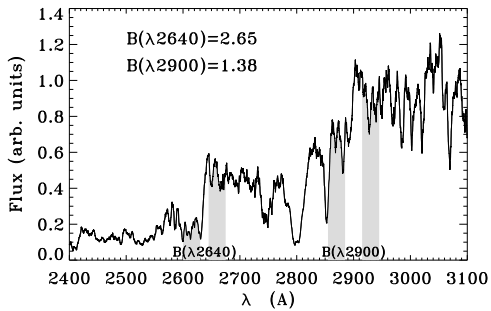}
\caption{
STIS E230M spectrum of 9 Comae, binned to 1~\AA .
The 2640 and 2900~\AA\ breaks of Spinrad et al.\ (1997)
are illustrated by the shaded panels (ratio of
the integrated flux longward of the break
to the flux shortward of the break). 
}
\end{figure}

\section{Analysis}

What are the constituent absorbers in the spectral breaks? Why do
they show the observed trends with spectral type? Can the observed spectral
breaks be reproduced by model spectra at the correct temperature, gravity,
and metallicity? Are the spectral breaks reliable indicators of age?
To help answer these questions, 
we computed the high-resolution spectra for comparison to STIS observations.
To generate the synthetic spectra, we applied our spectral synthesis
program, \textsc{synspec} (Hubeny 1988\markcite{H88}; 
Hubeny, Lanz \& Jeffery 1994\markcite{H94})
to fully line-blanketed model atmospheres provided by 
Kurucz (1993\markcite{K93}).

For this application, we assumed
LTE for all ions, and we used line opacities as derived from the Kurucz
(1993\markcite{K93}) line list. We used all lines (those originating between
predicted as well as measured levels) for 
\ion{Fe}{1}, \ion{Fe}{2}, and \ion{Ni}{2}, but only
lines originating between measured levels for other metals. We made extensive
tests to verify that this simplification does not lead to any appreciable
errors. We included all relevant molecular lines from 
Kurucz' (1993\markcite{K93}) 
data set. We calculated continuum opacities using energy levels and
cross-sections from the TOPbase interface 
(Cunto et al.\ 1993\markcite{C93}) to
the Opacity Project database. To ensure that continuum edges were properly
placed, we shifted the theoretical energy levels in TOPbase to
match the laboratory measurements of Martin et al.\ (1995\markcite{M95}). 

Figure 4 compares the model spectrum of 9 Comae to a sample segment 
(2870--2880~\AA) of the STIS spectrum, shown at full resolution.
While most features are readily identified, often there is an
observed spectral feature that is not reproduced by the model, and vice
versa.
%; i.e.\ there is a spectral feature in the model spectrum that does not
%occur in the observed spectrum. 
We attribute this lack of correspondence to
wavelength errors in the line list, since it appears to occur randomly and
in both directions. In general, though, the match of model to observations
is good enough that we can have confidence in the line identifications.

\begin{figure*}[!t]
\plotfiddle{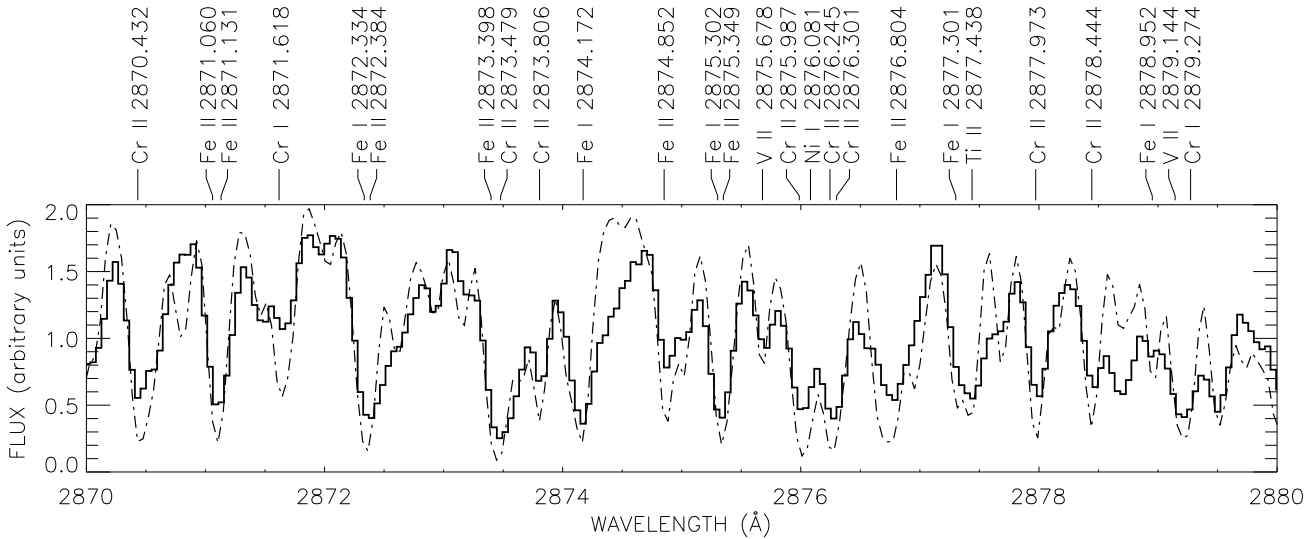}{6in}{90}{80}{80}{305}{250}
\vskip -3.7in
\caption{
Sample segment of the STIS E230M spectrum of 9 Comae (bold) compared
with \textsc{synspec} model (dot-dash) 
for the parameters of the star (Edvardsson et al.\ 1993).
}
\end{figure*}

What causes the spectral breaks?  Morton et al.\ (1977\markcite{M77}) 
identified the principal contributors to the features in this region, using 
{\it Copernicus} data.  Our own models show that the
spectral interval incorporated in the 2640~\AA\ break has only one weak 
\ion{Fe}{1} absorption edge, while the 2900~\AA\ break contains only
minor \ion{Mg}{1} photoionization resonance features. Thus, the breaks are
not a measure of the strength of continuum edges. Rather, they measure the
chance agglomerations of lines. In the case of the 2640~\AA\ break, the
absorption trough is dominated by lines belonging to the UV1 resonance
multiplet of \ion{Fe}{2}, while longward of the break, the absorption is
produced by subordinate \ion{Fe}{2} lines and lines from other elements.
The sensitivity of B($\lambda$2640) to temperature is an excitation effect in
\ion{Fe}{2}, the dominant ion of Fe. At increasing temperatures, the
excited states of \ion{Fe}{2} become populated at the expense of the
ground-state population.

Figure 5 compares our \textsc{synspec} results 
(assuming the abundances of Edvardsson et al.\ 1993) to the observed UV
breaks in the spectrum of 9 Comae. The model reproduces the two spectral
breaks for the temperature, gravity, and elemental abundances of 9 Comae,
implying that the model is essentially correct. Furthermore, it faithfully
follows the observed trends of B($\lambda 2640$) with temperature (Fanelli 
et al.\ 1992\markcite{F92}; Spinrad et al.\ 1997\markcite{S97}). 
The \textsc{synspec} models also reveal
a significant sensitivity to gravity, which might explain the dispersion
in the observed measurements of B($\lambda 2640$) in luminosity-class V and 
IV stars. In contrast to the general agreement between the 
\textsc{synspec} models 
and STIS observation, Kurucz' (1993\markcite{K93}) 
spectral energy distributions (SEDs) systematically 
overestimate the 2640~\AA\ break and therefore should not be used to 
infer age from this diagnostic. Both the \textsc{synspec} models 
and Kurucz SEDs
match the 2900~\AA\ break in the spectrum of 9 Comae as shown.
However, compared to observations presented by 
Fanelli et al.\ (1992\markcite{F92}) and Spinrad et al.\ (1997\markcite{S97})
both fail at temperatures below about 6000~K.

\begin{figure}[h]
\plotone{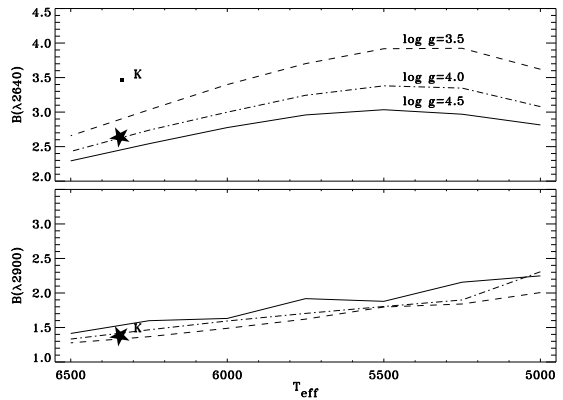}
\caption{
Comparison of model UV spectral breaks (with the appropriate 
Edvardsson et al.\ 1993 abundances) to the observed breaks 
(star symbol) in the spectrum of 9 Comae.  Kurucz' predictions for 9 Comae
are also shown (square with ``K'' symbol), assuming [M/H]=+0.3.
}
\end{figure}

We calculated the UV spectra of stellar systems described by 
Yi \& Demarque's (1997\markcite{Y97})
isochrones for metallicities of one and two times solar.  
These isochrones are similar to the Yale 
(Green, Demarque, \& King 1987\markcite{G87}) isochrones, but
with improved opacities.  Figure 6 
compares the resulting computed UV spectral breaks to the observed ones 
in the spectrum of 53W091.  B($\lambda$2640) yields an age of 1 to 2 Gyr 
depending on the assumed metallicity. B($\lambda$2900) yields
a much older age ($\geq$5 Gyr) for 53W091, but this diagnostic, while 
useful in empirical studies, is not to be trusted here, 
since our \textsc{synspec} models (and Kurucz' SEDs) do not match the
observed values at temperatures lower than 6000 K.  We will investigate this 
discrepancy once we obtain STIS UV spectra of cooler stars.  

\begin{figure}[h]
\plotone{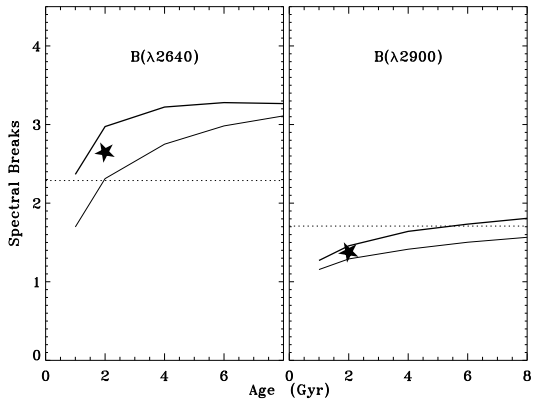}
\caption{
Model UV spectral breaks of stellar systems as a function of age.
[M/H]~=~+0.3 models are shown in bold, [M/H]~=~0 as thin lines. 
The measured breaks in 53W091 (dotted line) and 9 Comae (star symbol) are
also plotted.
}
\end{figure}

\section{Conclusions and Future Work}

In summary, the STIS spectrum of 9 Comae has verified our spectral model
of a late F-type star near the main sequence turnoff;
the 2640~\AA break yields the same temperature and age as its location
on the HR diagram.  With this observational
validation, we have computed the age-sensitive 2640~\AA\ spectral break for 
stellar systems of various ages. Our derived age ($\lesssim$ 2 Gyr) for the 
youngest stars in the
early-type galaxy, 53W091, is considerably lower than estimated by 
Spinrad et al.\ (1997\markcite{S97}); 
consequently, it relaxes the constraints of cosmological parameters that
were implied in that work.

This Early Release Observation suggests that future STIS observations will
help to fine-tune stellar atmospheric models;
this in turn will provide a direct 
conversion from UV spectral characteristics to the fundamental parameters of 
a star or stellar system. 
However, it also identifies important spectral 
modelling issues (e.g., the importance of chromospheric 
emission to age-sensitive
features; the validity of assuming LTE; and most importantly, the adequacy of 
the atomic and molecular data) as well evolutionary modelling issues (e.g., 
the importance of convective overshooting). These issues can also
best be addressed through future STIS observations.

We gratefully acknowledge the advice and criticisms of the referee, Hyron
Spinrad.  This study has made extensive use of SIMBAD.

\end{document}